\documentclass[twocolumn,showpacs,preprintnumbers,amsmath,amssymb,floatfix]{revtex4}

\usepackage{amsmath,amsfonts,graphics,graphicx,epsfig,amssymb}
\usepackage{dcolumn}
\usepackage{bm}

\newcommand\ds{D_s^+ }
\newcommand\fd{f_{D_s^+} }
\newcommand\nuh{\nu_h \to  \gamma \nu}
\newcommand\dsmh{D_s^+ \to  \mu^+ \nu_h}
\newcommand\dsmm{D_s^+ \to  \mu^+ \nu_\mu}
\newcommand\dstt{D_s^+ \to  \tau^+ \nu_\tau}

\newcommand\mix{|U_{\mu h}|^2}
\newcommand\Br{{\rm Br}}
\renewcommand\l{\left(}
\renewcommand\r{\right)}

\def\address{\@ifstar{\address@star}%
  {\@ifnextchar[{\address@optarg}{\address@noptarg}}}

\begin{document}

\author{S.N.~Gninenko\footnote{ Sergei.Gninenko\char 64 cern.ch}}
\author{D.S.~Gorbunov\footnote{Dmitry.Gorbunov@cern.ch}}

\affiliation{Institute for Nuclear Research of the Russian Academy of Sciences, Moscow 117312}


\title{
The MiniBooNE anomaly, the decay $\dsmm$ and heavy sterile neutrino}

\date{\today}

\begin{abstract}
It has been recently suggested that the anomalous excess of low-energy 
electron-like events observed by the MiniBooNE experiment, could be
explained by the radiative decay of a heavy sterile neutrino $\nu_h$ 
of the mass around 500 MeV with a muonic mixing strength in
the range $|U_{\mu h}|^2 \simeq (1-4)\times 10^{-3}$.  If such $\nu_h$
exists its admixtures in the decay $\dsmm$ would result in the decay
$\dsmh$ with the branching fraction $\Br(\dsmh)\simeq (1.2-5.5)\times
10^{-4}$, which is in the experimentally accessible range.
Interestingly, the existence of the $\dsmh$ decay at this level may
also  explain why the currently measured decay rate of
$\dsmm$ is slightly higher than the predicted one. This enhances
motivation for a sensitive search for this decay mode and makes it
interesting and complementary to neutrino experiments
probing sterile-active neutrino mixing.  Considering, as an example
the CLEO-c experiment, we suggest to perform a search for the decay
$\dsmh$ with the analysis of existing data. The discrepancy between
the measurements and theoretical description of the decay $\dstt$ is
also discussed in brief.
 
\end{abstract}
\maketitle


\section{Introduction}

The MiniBooNE collaboration, which studies the interactions of neutrinos from
the  $\pi^{+}$ decays in flight at FNAL,
has observed an excess of low energy electron-like events
   in the energy distribution of charge-current quasi-elastic
electron neutrino events \cite{mb1}. 
This anomaly
has been  recently further confirmed 
with larger statistics \cite{mb2}. 
As the collaboration has not yet clarified
the origin of the excess, several models involving new physics   were
considered to explain the discrepancy, see e.g. \cite{mb2} and references 
therein.

It is well known,  that the neutrino weak flavor eigenstates
($\nu_e,~\nu_{\mu},~\nu_{\tau},...$) need not coincide with the mass
eigenstates ($\nu_1,~\nu_2,~\nu_3,~\nu_4...$), but would, in general,
be related through a unitary transformation.   Such a generalized mixing:
\begin{equation}
\nu_l= \sum_i U_{li} \nu_i;~~~l=e,\mu,\tau,...,~i=1,2,3,4,...
\label{mixing}
\end{equation}
results in neutrino oscillations when the mass differences are small,
and in
decays of heavy neutrinos when the mass differences are large. 
\begin{figure}[htb!]
    \includegraphics[width=\columnwidth]{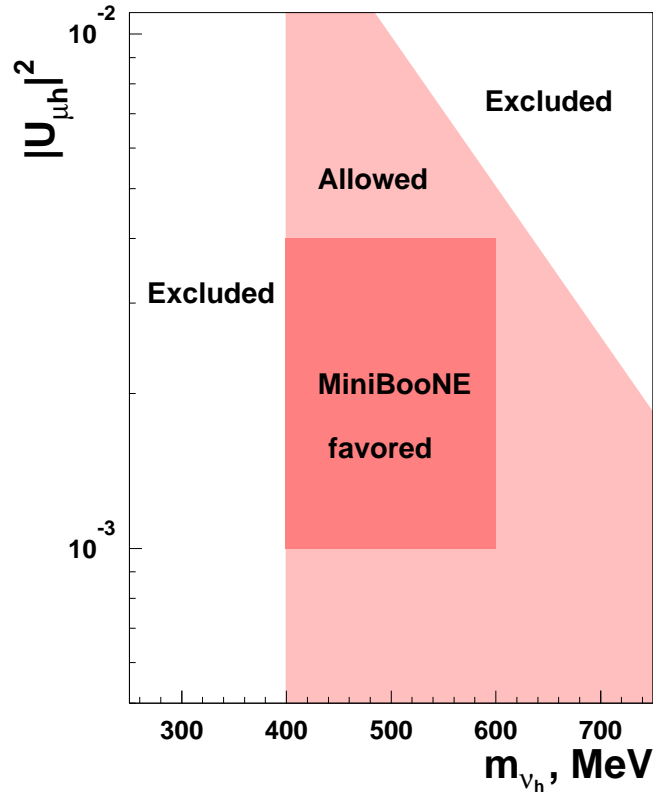}
     \caption{The shaded area is the experimentally allowed region of 
 the mixing strength 
$\mix$ 
calculated for  $\mu_{tr} = 10^{-9} \mu_B$
in  the model of Ref.~\cite{sng}. The rectangular area represents
the region in parameter space  
favorable  for the suggested in Ref.~\cite{sng} 
explanation of the MiniBooNE anomaly. 
\label{mix}}
\end{figure}
In the recent work \cite{sng} it has been shown that the MiniBooNe
excess could be explained by the production of a sterile neutrinos, 
$\nu_h$'s, of the mass around $\simeq$ 500 MeV, which, being created by
mixing in $\nu_\mu$ neutral-current interactions,  decay  
(dominantly) into photons and light neutrinos in the MiniBooNE
detector target. Such kind of $\nu_h$ could arise in many interesting
extensions of the Standard Model (SM), such as GUTs, Superstring
inspired models, Left-Right Symmetric models, and others.  It can decay
radiatively into $\nu \gamma$, if e.g. there is a non-zero transition
magnetic moment ($\mu_{tr}$) between the $\nu_h$ and active neutrino
$\nu$ \cite{moh}.  The required mixing strength
\begin{equation}
|U_{\mu h}|^2 \simeq (1-4)\times  10^{-3}
\label{ran1}
\end{equation} 
was found to be consistent with existing experimental data for 
$\mu_{tr}\simeq (1-6)\times 10^{-9} \mu_B$ (here $\mu_B$ is the Bohr
magneton) \cite{sng}.  For illustration, experimentally allowed region
of the mixing strength $\mix$ in the $\nu_h$ mass range
around 500 MeV is shown in Fig.~\ref{mix} for $\mu_{tr}
= 10^{-9} \mu_B$ together with the parameter region favorable for
the explanation of the MiniBooNE anomaly. It worth to mention that the
model \cite{sng} is also consistent with  the absence of a significant 
low-energy excess in MiniBooNe antineutrino data \cite{antinu}.

In this letter we put forward an idea that sterile-active neutrino
mixing in the allowed range 
shown in Fig.~\ref{mix}  
could be tested by searching for 
the admixtures of $\nu_h$ in the decay $\dsmm$. In addition
we point out that the present discrepancy of about 3$\sigma$ 
between the 
measured and predicted 
decay rate of $\dsmm$ could be explained by the unrecognized
contribution from the decay $\dsmh$.


\section{The decay $\dsmm$ and heavy neutrino}

If the $\nu_h$ exists, it could be a component of $\nu_\mu$,
and as follows from Eq.~\eqref{mixing}, would be produced by any source
of $\nu_\mu$ according to the mixing $|U_{\mu h}|^2$ and
kinematic constraints. In particular, $\nu_h$ could be produced in any
leptonic and semileptonic decays of sufficiently heavy mesons and
baryons. For the interesting mass range $m_{\nu_h}\simeq400-600$~MeV
the most promising process is the leptonic decay $\dsmm$.

In the SM, $D_s$ meson decays leptonically via annihilation of the $c$
and $\overline{s}$ quarks through a virtual $W^+$.  The decay rate of
this process is given by
\begin{equation}
\Gamma(D_s^+ \to l^+ \nu) = \frac{G_F^2}{8 \pi}f^2_{D_s^+} m_l^2
M_{D_s^+} \!\!
\l \!\!1-\frac{m_l^2}{M_{D_s^+}^2} \!\r^{\!\!2} \!\!|V_{cs}|^2,
\label{rate1}
\end{equation}
where the $M_{D_s^+}$ is the $D_s^+$ meson mass, 
$m_l$ is the mass of the charged 
lepton, $\fd$ is the decay constant, $G_F$ is the Fermi 
constant, and $V_{cs}$ is a Cabibbo-Kobayashi-Maskawa matrix element 
which value equals 0.97334 \cite{pdg}. The decay rate \eqref{rate1} 
is suppressed by the lepton mass squared, since the
very leptonic decay is due to chirality-flip. 

\begin{figure}[!htb]
\includegraphics[width=\columnwidth]{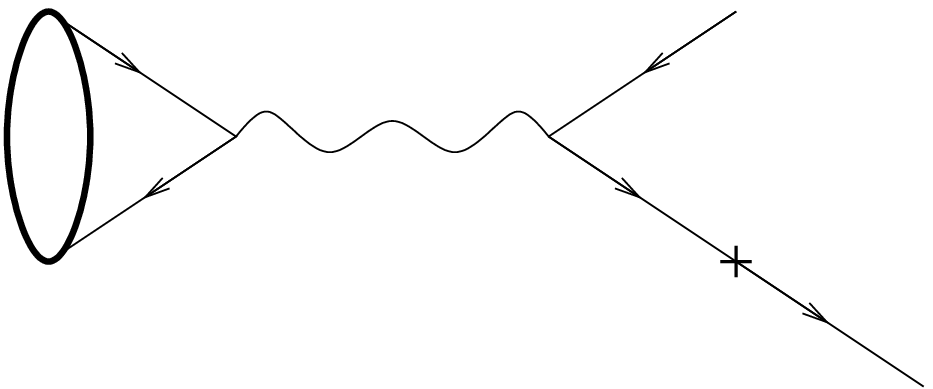}
\begin{picture}(0,0)
{\normalsize
\put(40,100){$\mu^+$}
\put(40,50){$\nu_\mu$}
\put(90,15){$\nu_h$}
\put(78,50){$U_{\mu h}$}
\put(-80,100){$c$}
\put(-80,50){$\bar s$}
\put(-20,90){$W^+$}
\put(-117,77){$D_s^+$}
}
\end{picture}
     \caption{ Schematic illustration of the  decay $\dsmh$.   
\label{diag}}
\end{figure}

The  mixing between the sterile neutrino and muon neutrino 
results in the  decay $\dsmh$, as illustrated  in Fig.~\ref{diag}. 
For the
interesting mass interval $m_{\nu_h}\simeq400-600$~MeV 
the chirality-flip is mostly due to sterile neutrino mass  
which results in 
\begin{equation}
\Gamma(D_s^+ \to \mu^+ \nu_h) \approx 
\Gamma(D_s^+ \to \mu^+ \nu_\mu)|U_{\mu h}|^2
\Bigl(\frac{m_{\nu_h}}{m_\mu}\Bigr)^2\;.
\label{rate3}
\end{equation} 
Using Eq.~\eqref{ran1} and taking into account the most precise
determinantion of the $\dsmm$ branching ratio $\Br(\dsmm)=\l 
0.565\pm0.045\pm 0.017\r \% $ \cite{alex}, 
we find that the branching fraction of $\dsmh$ is in 
the experimentally accessible range:
\begin{equation}
\Br(\dsmh)\approx 
(1.2-5.5)\times 10^{-4} \l \frac{m_{\nu_h}}{500~{\rm MeV}}\r^2\;.
\label{br1}
\end{equation}  

\section{Direct experimental search for the decay  $\dsmh$ }

Consider now, as an example, the CLEO-c experiment,  where the search for the 
decay $\dsmh$ could be performed. 
In this experiment  
several of the most precise measurements of properties of $\ds$ mesons
have been performed  by using the CLEO-c detector at 
CESR \cite{cleodet}. Recently, the CLEO 
 collaboration studying  the process $e^+ e^-
 \to D_s^- D_s^{*+}, D^{*-}_s D_s^+ $   has reported on measurements
of the decay constant $f_{D_s^+}$ of $D_s$   mesons to a precision of a few \%
\cite{alex,ony}, see also \cite{cleoexp}. 

The detector is well equipped to identify and measure the momenta/energy
  and directions of charged particles and  photons. 
The experiment was performed at a centre-of-mass energy of 4170~MeV, 
where the cross section of a charmed meson pair production 
is relatively large. This allowed to fully reconstruct the $D_s^-$ as a 'tag'
and study the leptonic decay properties of the other through the decay chains 
\begin{equation}
\begin{split}
e^+e^- \to D_s^- D^{*+}_s  \to D_s^- \gamma D^+_s  \to D_s^- \gamma \mu^+  \nu  \\
e^+e^- \to D^{*-}_s D_s^+ \to \gamma D^-_s D_s^+ \to \gamma D_s^- \mu^+  \nu 
\end{split}
\label{chain1}
\end{equation}
The decays $\dsmm$ were identified by selecting the events with a
single missing massless neutrino, for which the missing mass-squared, $MM^2$,
evaluated by taking into account the reconstructed $\mu^+, ~D_s^-$
and $\gamma$ should peak at zero. The $MM^2$ is calculated as
\begin{equation}
\begin{split}
MM^2=(E_{CM}-E_\mu - E_\gamma-E_{D_s})^2 \hskip 2cm \mbox{} \hfill \\ 
\mbox{} \hfill -(p_{CM}- p_\mu - p_\gamma-p_{D_s})^2
\end{split}
\label{mm}
\end{equation}
where $E_{CM}$ and $p_{CM}$ are the centre-of-mass energy and 3-momentum,
$E_{D_s}$ and $p_{D_s}$ are the energy and 3-momentum of the fully
reconstructed $D_s^-$ tag, $E_{\gamma}$ and $p_{\gamma}$ are the
energy and 3-momentum 
of the photon and $E_{\mu}$ and $p_{\mu}$ are the energy
and 3-momentum of the muon \cite{alex}.

Similarly to this approach, the basic idea of probing the model under
 discussion is to search for a peak corresponding to the value
 $m_{\nu_h}^2$ in the $MM^2$ distribution. It should
 be calculated taking into account measured properties of observed
 $\mu,~\ds$ and photon from the decay chains:
\begin{equation}
\begin{split}
e^+e^-\to D_s^- D^{*+}_s \to D_s^- \gamma D^+_s\to D_s^- \gamma  \mu^+
\nu_h \to  
\\
\hfill D_s^- \gamma  \mu^+ \gamma_h \nu  \\
e^+e^-\to D^{*-}_s D_s^+\to \gamma  D_s^- D^+_s\to \gamma D_s^- \mu^+
\nu_h \to  
 \\
\hfill \gamma D_s^- \mu^+ \gamma_h \nu 
\end{split}
\label{chain2}
\end{equation}
where $\gamma_h$ denotes a photon from the dominant decay mode $\nuh$ of
sterile neutrino.  In the largest part of the ($|U_{\mu h}|^2, \mu_{tr}$) 
parameter space favored by MiniBooNE, the $\nu_h$ is expected to 
be a short-lived particle with the lifetime less than $ 10^{-9}$~s \cite{sng}.
Then, its  decay length is significantly less than the
 radius of the CLEO detector (95 cm), and most of the $\nuh$ decays would 
occur inside the CLEO-c detector fiducial volume in the vicinity of
 the primary vertex.  

The experimental signature of the decay $\nuh$ is a peak in the mass range
$0.16-0.36$~GeV$^2$ of the distribution of \eqref{mm}.
Using Eqs.\eqref{ran1},\eqref{rate3} and the total number of $(235.5\pm13.8)$ $\mu^+ \nu_\mu$
events observed by CLEO-c with 600 pb$^{-1}$ \cite{alex},
$(5-22)\times \l \frac{m_{\nu_h}}{500~{\rm MeV}}\r^2$  
events are expected to be found at the peak.
To obtain  the correct $MM^2$ value,  
 the photon from the decay $\nuh$ should not be used in calculations of 
 Eq.~(\ref{mm}).
For the energy greater than threshold, $E_{\gamma_h}> 300$ MeV, the 
$\gamma_h$ could  be identified as an extra photon in the event
candidate for the decay chains (\ref{chain1}).

Finally, note that a search for the decay mode $\nu_h \to \mu \pi$  is also 
of a special interest. Although this decay is sub-dominant, for the mixing as 
large as in Eq.~(\ref{ran1}) its branching fraction could be 
of the order of few \% \cite{sng}, and 
a few events could  be observed in the CLEO-c experiment for 
the 600 pb$^{-1}$ of data. The experimental signature of 
the event $\nu_h \to \mu \pi$ would be two charged tracks originated from a 
 common vertex  displaced from the primary vertex. Since there is no neutrino
in the final state, it is possible to reconstruct the invariant mass of the
heavy sterile neutrino, that would manifest itself as a peak 
 in the range 0.16-0.36 GeV$^2$ of  the invariant mass squared. 
An observation of a few $\mu \pi$-events with the same invariant 
mass  would provide an excellent cross-check of the model.


\section{The $D_s^+\to \mu^+  \nu_\mu, \tau^+ \nu_\tau$ decays puzzle}

Interestingly, the above discussions might be relevant to the
discrepancy of about 3$\sigma$ between the measured and predicted
rates of $\dsmm$, see e.g. Refs.~\cite{alex,ony} and discussion
therein. Presently, in spite of the substantial theoretical and
experimental efforts, the decay rates of $\dsmm$ 
measured by
the CLEO-c \cite{alex}, Belle \cite{bel}, and BABAR \cite{bab}
experiments are found to be slightly higher than the predicted one,
most accurately calculated in the framework of lattice QCD
\cite{follana}.  Taking into account the most precise results on the
$\dsmm$ decay rate from these experiments one arrives at 
\begin{equation}
\label{disc}
\begin{split}
\frac{\Gamma^{exp}(\dsmm)-\Gamma^{th}(\dsmm)}{\Gamma^{exp}(\dsmm)}
\hfill \\ \hfill = 0.166\pm0.060 
\end{split}
\end{equation}
where $\Gamma^{exp}(\dsmm)$ and 
$\Gamma^{th}(\dsmm)$ are the average measured 
and predicted values for the $\dsmm$ decay rate, respectively, 
with the statistical and 
systematical uncertainties combined in quadrature. Thus, 
the ratio of Eq.~\eqref{disc} differs from zero by about 2.8$\sigma$
standard deviations.

Various models of new physics giving additional contribution 
to the rate of $\dsmm$ 
have been investigated in order to resolve the discrepancy, 
see e.g. \cite{dob,Dorsner:2009cu}. 
We propose here that the reason of why the experimental rate of $\dsmm$ 
is higher than the theoretical expectations may be due  
the contribution from the decay $\dsmh$. 

Consider again, as an example, search for $\dsmm$ events in the CLEO
experiment \cite{alex}.  If the $\gamma_h$ from the decay chain
(\ref{chain2}) is used, the calculated $MM^2$ should peak at zero
regardless of wether or not the $\gamma_h$ is produced in the direct
$\ds$ decay. This is valid under assumption that the $\nu$ from the
$\nuh$ decay is light. Thus, the events  (\ref{chain2}) may be
accepted and contribute to the number of the $\dsmm$ signal events.
Using Eqs.~(\ref{rate1},\ref{rate3},\ref{disc}) one finds
that in order to explain the discrepancy the 
branching fraction of decay mode to sterile neutrinos of masses
$m_{\nu_h}\simeq 400-600$~MeV should be within the range 
\begin{equation}
\Br(\dsmh)=\frac{(9.85\pm 4.16)\times 10^{-4}}{k} \;,
\label{br2}
\end{equation} 
where factor $k=\frac{P(\dsmh)}{P(\dsmm)}$ is the ratio
of the overall probabilities for the events (\ref{chain2}) and
(\ref{chain1}) 
to pass selection criteria in the analysis of $\dsmm$ events
in \cite{alex}. In this search, for the selection of $\dsmm$ candidates
it was required that there should be no additional photon, not
associated with the tag, detected in the ECAL with energy greater than
300 MeV (photon veto) \cite{alex}.  The fraction of $\gamma_h$ from
(\ref{chain2}) that would pass this veto cut is roughly estimated to
be $k \simeq 40\% $. This results in
\begin{equation}
\Br(\dsmh)=(24.6\pm 10.4)\times 10^{-4}
\;.
\label{br3}
\end{equation}
The corresponding mixing strength is (c.f. \eqref{br1} and
\eqref{ran1}) 
\begin{equation}
\mix = (25.2\pm 10.7)\times 10^{-3}
\l \frac{500~{\rm MeV}}{m_{\nu_h}}\r^2 \;.
\label{ran3}
\end{equation}
The explanation of the MiniBooNE anomaly implies somewhat smaller
contribution to the decay  $\dsmm$.  
However, the regions for the obtained  mixing strength of Eq.~\eqref{ran3} 
and branching fraction of Eq.~\eqref{br3} overlap at the level of less than 2
$\sigma$  with the ranges of Eqs.~\eqref{ran1} and \eqref{br1}, respectively, 
required to explain
the MiniBooNE anomaly, see also Fig.~\ref{mix}. This  enhances 
motivation for a sensitive search of the decay $\dsmh$.

In close analogy with the case of $\ds$, the mixing \eqref{ran1} can
also be probed with study of the leptonic decay rates of the
$D^+$-meson. In particular, our model implies the quite similar to
\eqref{rate3} contribution to the total muonic decay rate of
 $D^+\to\mu^+ \nu$. However, the achieved accuracy in
measurement of this decay rate \cite{pdg} is worse than that for the
$D_s$-meson, hence the suggested new contribution is unrecognizable
yet.

It should be mentioned that the best precision in measurements of the decay
rate  of  $\dstt$ has aslo been achieved at CLEO~\cite{ony}, and the
result is 
consistent with theoretical predictions. However from other
measurements at CLEO~\cite{alex}, 
based on study of another decay mode of outgoing
$\tau$-lepton, the obtained decay
rate  of  $\dstt$ is significantly higher than the predicted one, so
that combined branching ratio deviates from the theoretical prediction 
at the level comparable 
to Eq.~\eqref{disc}. Thus, one may speculate that this  
discrepancy is due to existence of an additional
sterile neutrino coupled to the tau neutrino.
Similar to above consideration 
we found that the mixing strength required to explain $\dstt$ discrepancy 
should be 
\begin{equation}
|U_{\tau h}|^2 = 0.16 \pm 0.09 
\label{ran4}
\end{equation}
for the mass range much below 190 MeV, where phase space suppression is
negligible. Note, that this result is consistent 
with direct experimental limits \cite{nomad, delphi}.

It is worth noting that heavy sterile neutrino of mass about 500~MeV
and mixing to muon neutrino as strong as
\eqref{ran1}, \eqref{ran3} can be searched in other
decays of charmed hadrons, beauty hadrons and $\tau$-leptons. Pure
leptonic modes are more promising, since heavy neutrino contribution
is enhanced with respect to that of the SM by a squared mass ratio of
sterile neutrinos and charged lepton.  This is not the case for other
decay modes which branching ratios are roughly the same as
corresponding decay to active muon neutrino multiplied by $|U_{\mu
h}|^2$, and somewhat suppressed due to reduced phase space volume.
By making use of general formulae for meson and $\tau$-lepton decays
to sterile neutrinos presented in Ref.~\cite{gs} and similar formulae
for baryons \cite{baryons} one can obtain more accurate estimates.

\section{Summary}

To summarize, we show that the 
 recently suggested  explanation~\cite{sng} of the MiniBooNE
 anomaly~\cite{mb1,mb2} can be probed by studying leptonic
 decays of charmed mesons. This study can be undertaken with already
 collected data. We speculate that the present 2.8 $\sigma$ 
discrepancy  between the 
measured and predicted rate of $\dsmm$ can be a hint at the presence
 of heavy sterile neutrino suggested in Ref.\cite{sng}. 

{\bf Acknowledgements}

We thank F.L. Bezrukov, N.V. Krasnikov, V.A. Matveev, V.A. Rubakov, and
   M.E. Shaposhnikov 
for useful discussions. This work of D.G. was partially
supported by the Russian Foundation for Basic Research (RFBR) grant
08-02-00473a, by the Russian Science
Support Foundation, by the grants 
of the President of the Russian Federation NS-1616.2008.2 and 
MK-1957.2008.2. S.G. has been supported by the RFBR 
grants 07-02-00256 and 08-02-91007. 
D.G. thanks for a hospitality LPPC ITPP EPFL where
a part of this work was done.	


\begin{thebibliography}{20}

\bibitem{mb1}
 A.~A.~Aguilar-Arevalo {\it et al.}  [The MiniBooNE Collaboration],
  Phys.\ Rev.\ Lett.\  {\bf 98}, 231801 (2007)
  [arXiv:0704.1500 [hep-ex]].

\bibitem{mb2}
A.~A.~Aguilar-Arevalo {\it et al.}  [MiniBooNE Collaboration],
  Phys.\ Rev.\ Lett.\  {\bf 102}, 101802 (2009)
  [arXiv:0812.2243 [hep-ex]].


\bibitem{sng}
S.~N.~Gninenko,
  arXiv:0902.3802 [hep-ph].

\bibitem{moh}
See for example, R.N. Mohapatra and P.B. Pal, "Massive Neutrinos in
Physics
and Astrophysics", World Scientific, Singapore, 1991;

\bibitem{antinu}
 A.~A.~Aguilar-Arevalo {\it et al.},
  arXiv:0904.1958 [hep-ex].

\bibitem{pdg}  C.~Amsler {\it et al.}  [Particle Data Group],
  Phys.\ Lett.\  B {\bf 667}, 1 (2008).


\bibitem{alex}
 J.~P.~Alexander {\it et al.}  [CLEO Collaboration],
  Phys.\ Rev.\  D {\bf 79}, 052001 (2009)
  [arXiv:0901.1216 [hep-ex]].


\bibitem{cleodet}
  D.~Peterson {\it et al.},
  Nucl.\ Instrum.\ Meth.\  A {\bf 478}, 142 (2002).

\bibitem{ony}
P.~U.~E.~Onyisi {\it et al.}  [CLEO Collaboration],
  Phys.\ Rev.\  D {\bf 79}, 052002 (2009)
  [arXiv:0901.1147 [hep-ex]].



\bibitem{cleoexp} 
  T.~K.~Pedlar {\it et al.}  [CLEO Collaboration],
  Phys. \ Rev. \ D {\bf 76}, 072002 (2007)
  [arXiv:0704.0437 [hep-ex]].





\bibitem{bel}
  K.~Abe {\it et al.}  [Belle Collaboration],
  Phys.\ Rev.\ Lett.\  {\bf 100}, 241801 (2008)
  [arXiv:0709.1340 [hep-ex]].

\bibitem{bab}
  B.~Aubert {\it et al.}  [BABAR Collaboration],
  Phys.\ Rev.\ Lett.\  {\bf 98}, 141801 (2007)
  [arXiv:hep-ex/0607094].

\bibitem{follana}
 E. Follana et al.,
  Phys.\ Rev.\ Lett.\  {\bf 100}, 062002 (2008)



\bibitem{dob}
 B.~A.~Dobrescu and A.~S.~Kronfeld,
  Phys.\ Rev.\ Lett.\  {\bf 100}, 241802 (2008)
  [arXiv:0803.0512 [hep-ph]].

\bibitem{Dorsner:2009cu}
  I.~Dorsner, S.~Fajfer, J.~F.~Kamenik and N.~Kosnik,
  arXiv:0906.5585 [hep-ph].


\bibitem{nomad}
  P.~Astier {\it et al.}  [NOMAD Collaboration],
  Phys.\ Lett.\  B {\bf 506}, 27 (2001)
  [arXiv:hep-ex/0101041].

\bibitem{delphi}
P.~Abreu {\it et al.}  [DELPHI Collaboration],
  Z.\ Phys.\  C {\bf 74}, 57 (1997)
  [Erratum-ibid.\  C {\bf 75}, 580 (1997)].

\bibitem{gs}
D.~Gorbunov and M.~Shaposhnikov,
  JHEP {\bf 0710}, 015 (2007)
  [arXiv:0705.1729 [hep-ph]].

\bibitem{baryons}
S.~Ramazanov,
  arXiv:0810.0660 [hep-ph].


\end{thebibliography}
\end{document}